\documentclass[a4paper,10pt]{article}
\usepackage{amsfonts}
\usepackage{amssymb}
\usepackage{pdfpages}
\usepackage{latexsym}
\usepackage{enumerate}
\usepackage{amsmath}
\usepackage{amsthm}
\usepackage{graphicx}
\usepackage{multirow}

\begin{document}

\title{Regularities and Discrepancies of Credit Default Swaps: \\
a Data Science approach through Benford's Law}
\author{Marcel Ausloos$^{1}$\thanks{%
Corresponding author.},$\;$ Rosella Castellano$^{2}$, $\;$Roy Cerqueti$^{3}$
}
\date{ $^{1}$ School of Management,University of Leicester, \\
University Road, Leicester LE1 7RH, UK\\
$e$-$mail$ $address$: ma683@le.ac.uk\\
\vskip0.5cm $^2$ University of Macerata, Department of Economics and Law,\\
via Crescimbeni 20, I-62100, Macerata, Italy \\
$e$-$mail$ $address$: rosella.castellano@unimc.it \\
\vskip0.5cm $^3$ University of Macerata, Department of Economics and Law,\\
via Crescimbeni 20, I-62100, Macerata, Italy \\
$e$-$mail$ $address$: roy.cerqueti@unimc.it}
\maketitle

\begin{abstract}
In this paper, we search whether the Benford's law is applicable to
monitor daily changes in sovereign Credit Default Swaps (CDS)
quotes, which are acknowledged to be complex systems of economic
content. This test is of paramount importance since the CDS of a
country proxy its health and probability to default, being
associated to an insurance against the event of its default. We fit
the Benford's law to the daily changes in sovereign CDS spreads for
13 European countries, -- both inside and outside the European Union
and European Monetary Union. Two different tenors for the sovereign
CDS contracts are considered: 5 yrs and 10 yrs, -- the former being
the reference and most liquid one. The time period under
investigation is 2008-2015 which includes the period of distress
caused by the European sovereign debt crisis. Moreover, (i) an
analysis over relevant sub-periods is carried out; (ii) several
insights are provided also by implementing the tracking of the
Benford's law over moving windows. The main test for checking the
conformance to Benford's law is -- as usual -- the $\chi ^{2}$ test,
whose values are presented and discussed for all cases. The analysis
is further completed by elaborations based on Chebyshev's distance
and Kullback and Leibler's divergence. The results highlight
differences by countries and tenors. In particular, these results
suggest that liquidity seems to be associated to higher levels of
distortion. Greece -- representing a peculiar case -- shows a very
different path with respect to the other European countries.
\end{abstract}



\section{Introduction}

Quantitative phenomena are characterized by numbers, whose values exhibit
specific features. For instance, it is clear that the distribution of
individuals' age in the world is not uniform, implying that it is more
probable to randomly extract a twenty-years old human than a hundred-years
old one. However, deviations from the uniform law have been observed for
large dataset also in not very intuitive contexts. In this respect, the
so-called Benford's law (BL), introduced by Newcomb (1881) and formalized by
Benford (1938), plays a prominent role. BL describes the regularity of the
frequency distribution of leading digits in some important large data set.
In particular, it states that the distribution of first digits is more
concentrated on smaller values, and the digit "one" has the highest
frequency. The story of BL is interestingly traced in Torres et al. (2007),
where both the theoretical advancements from the original formulation and a
list of its applications are provided. In brief, the assertion of the BL is
that, once taken a large set of numbers, the first digits distribution
follows a logarithmic law:
\begin{equation}
P(d)=log_{10}(1+\frac{1}{d}),\qquad d=1,2,\dots ,9,  \label{BL}
\end{equation}%
where $P(d)$ is the probability that some numbers from the data set have
their first digit equal to $d$; $log_{10}$ being the logarithm in base 10.
For the reader's convenience, in Table 1 we show the frequencies of the
first digit as given by the BL. \newline
The applications of BL range in a wide set of fields. A list of recent
contributions (published after the list of Torres et al.) should e.g.
include T{\"{o}}dter (2009), Bazzani et al. (2010), Rauch et al. (2011), De
and Sen (2011), Mir (2012, 2014), Mir et al. (2014), Ausloos and Clippe
(2012), Ausloos et al. (2015). In the specific context of financial
applications, BL violations might be meaningfully associated to data
misalignments (as much demonstrated by Nigrini 1992, 1996) and toxicity of
the markets, but also to other cases (Ausloos et al. 2015). Some papers are
particularly relevant here. Varian (1972) tested -- through BL -- the
presence of anomalies over a set of land planning data for 777 tracts,
roughly corresponding to census tracts in the area of San Francisco Bay:
input data (in year 1967) and forecasts (for years between 1970 and 1990).
He found surprisingly, yet somewhat expected, agreement (not his own words)
with BL.\newline
Nigrini (1992, 1996) assessed mistakes over a large collection of accounting
data, and explained them by invoking frauds or, simply, misprints in the
data reporting process (see also Nigrini and Mittermeir, 1997). It is also
worth noting the contributions of Ley (1996) who studied the stock market by
using Benford's Law (BL); Realdon (2008) who applied the BL and linked the
CDS market to the stock market, and Carrera (2015) who used BL to analyze
numerical patterns in exchange rates in order to verify whether they appear
to have been subject to some degree of policy management. \newline
In line with Abrantes et al. (2011), Judge and Schechter (2009) and Varian
(1972), we use the BL to verify the \textit{quality} and \textit{credibility}
of CDS data. This is a key issue since CDS are traded in decentralized Over
The Counter (OTC) markets which are often pictured as opaque and with little
information about pricing mechanisms, price settlement and trading volumes.
Moreover, we scientifically move from the perspective of CDS as complex
systems (see e.g. Giansante et al., 2012 and Kim and Jung, 2014, where the
CDS volatility is studied by adopting a network approach; supporting
arguments on the complex nature of the CDS can be also found in Burns et
al., 2013). In so doing, we are also in line with the literature stating
that BL provides an informative content mainly when the data generating
process is a complex system (see e.g. Li et al., 2015). \newline
Here, we adopt a data science perspective, aiming at deciphering the
presence of anomalies in the daily changes of sovereign CDS quotes. There
are three main reasons why we chose the sovereign CDS: first, before the
recent crisis, CDS were often lauded as derivative instruments that could
stabilize the financial system as a whole because of their portentous risk
transferring and risk signaling abilities. Secondly, the CDS market has
grown tremendously over the last decade and is currently an integral part of
the financial system. Thirdly, there is some anecdotal evidence that major
banks manipulated CDS prices. For example, it is worth citing the case of
Reuters reporting that big U.S. banks face CDS manipulation lawsuit\footnote{%
http://mobile.reuters.com/article/idUSL1N0R51TS20140904?irpc=932} and the
one of US Senator Carl Levin that has accused Goldman Sachs to manipulate
CDS quotes\footnote{%
http://blogs.reuters.com/felix-salmon/2010/12/10/annals-of-cds-manipulation-goldman-sachs-edition/%
}. \newline
Thus, the up-to-date relevance of the present investigation can be found in
the growing importance of the role played by sovereign CDS in the globalized
financial market (Longstaff et al., 2011). \newline
Indeed, after the financial crisis started in 2007 in the United States,
financial markets have found a new concern since the beginning of 2010: the
levels of deficits and public debts. All developed countries, even the
largest, are suspected of being able to default on their debt. Rating
agencies, bankers and investment funds have begun to worry about the
sustainability of public finances and required countries to reduce their
debt by cutting government spending, especially social spending. In this
context, it can be observed that the recent development of CDS on the debt
of developed countries has launched CDS as tools for estimating the
probability of default of the States, although their quotes are determined
on a opaque and poorly regulated market (Jarrow, 2011). As a consequence,
the scientific interest for sovereign CDS mainly focuses on a time span
which includes the key dates characterizing the financial crisis which is
still experienced. Indeed, a considerable number of sovereign CDS contracts
were signed in the early stages of the crisis (2008-2010), particularly on
the wake of the turbulence following the failure of Lehman Brothers.
Pre-crisis periods are generally associated to low volumes of these
products. Consequently, the limited meaningfulness of the analysis of their
paths allows us to shorten such considerations in this paper.

A broad range of literature analyzes the CDS market (Bao et al.,
2011; Castellano and D'Ecclesia, 2013; Castellano and Scaccia, 2014,
only to mention the most recent ones) and, especially in the last
six years, due to the European sovereign debt crisis, the scientific
literature has placed more emphasis on sovereign CDS. Many of the
recent studies have found that CDS quotes on debt securities
increase substantially before financial crises become full-blown.
However, it is worth to mention that, beyond its signaling power,
the CDS market still poses many questions in relation to its full
transparency and its capacity to spread of market disturbances --
mostly on the downside -- from one country to the other. For
instance, Afonso et al. (2012) use EU sovereign CDS spreads to carry
out an analysis on the reaction of spreads to announcements from
rating agencies. They show that quotes are sensitive to
announcements, and that spillover effects from lower to higher rated
countries among EMU countries, together with persistence effects,
can be observed. Longstaff et al. (2011) show that sovereign CDS
quotes are driven more by global market factors, risk premiums, and
investment flows than by country-specific macro-economic
fundamentals. In other words, their analysis supports a view of the
sovereign CDS market in which investors play a predominant role.
\newline
Some other studies highlight that the price discovery mechanism and the
knowledge of actual net positions of financial institutions are necessary to
ensure the transparency of the CDS market (Cont and Kokholm, 2014).
\begin{table}[tbp]
\begin{center}
\begin{tabular}{|l||l|l|l|l|l|l|l|l|l|}
\hline
\textbf{First digit} & 1 & 2 & 3 & 4 & 5 & 6 & 7 & 8 & 9 \\ \hline
\textbf{Frequency} & 0.301 & 0.176 & 0.125 & 0.097 & 0.079 & 0.067 & 0.058 &
0.051 & 0.046 \\ \hline
\end{tabular}%
\end{center}
\caption{Frequencies of the first digit in a set of data -- values ranging
from 1 to 9, see the first row -- according to BL.}
\end{table}

\section{Paper content}

The dataset considered in this paper is given by Thomson Reuters Composite
Sovereign CDS spreads for 13 European countries on a daily basis. Countries
are distributed in four main groups: a) the \textit{core economies} - France
Germany, United Kingdom; b) the \textit{most worrying economies} - Ireland,
Italy, Portugal and Spain; c) the \textit{Eastern economies} - Croatia,
Czech Republic, Poland, Romania - and Turkey; d) Greece. Spreads are
provided for two different tenors: 5 and 10 years. The period under
investigation is August 2008 - April 2015. Such a period includes the Lehman
Brother's bankruptcy and the sovereign debt crisis of developed countries,
which are associated to increasing level of signed sovereign CDS contracts
and to the interest in the analysis of their paths.\textbf{\ }The resulting
time series gives then an amount of 42,000 spread observations. \newline
The conformance between daily changes in CDS spreads and BL has been checked
by performing two different typologies of procedures. First, for both tenors
(5 and 10 years) a $\chi ^{2}$ test has been implemented over the entire
period and over four noticeable sub-periods : (i) August 8, 2008 - April 25,
2015; (ii) August 8, 2008 - January 1, 2010; (iii) January 1, 2010 - October
31, 2013; (iv) November 1, 2013 - April 25, 2015; (v) January 1, 2010 -
April 25, 2015. Such sub-periods were chosen to highlight the time span of
the European sovereign debt crisis and the one that preceded it. Second, for
the reference contract (5 years), a BL conformity track has been implemented
by computing three statistical indicators over moving windows of 120 days
length, with 60 days overlap: a $\chi ^{2}$ test, the Chebyshev's distance
and the Kullback and Leibler's divergence. \newline
It will be shown that several insights can be derived from the analysis:
first of all, there is much evidence of conformity with BL, but not in all
cases. In particular, the BL validity hypothesis can be accepted for a wide
set of countries, but only in the 10 yrs tenor case. Differently, BL is
rather systematically violated in the case of most liquid products, i.e. in
the 5 yrs tenor case. The core economies show remarkable discrepancies with
respect to BL, suggesting a major deviation of the daily changes sovereign
CDS spreads. It seems also that such violations are more evident after 2010,
and this leads to conjecture the presence of a relationship between the
financial distress and non conformity of CDS data to BL. Greece can be
viewed as a very special case, in that it follows a very different path with
respect to the other European countries. \newline
Of course, we cannot exclude the presence of scale effects biasing the
obtained results. In fact, CDS quotes can be very shallow since net and
gross CDS volumes for all the tenors represent typically a small fraction of
the outstanding government debt. In this respect, the most liquid one --
i.e. the 5yrs-CDS -- represents an even smaller fraction of the total
outstanding debt. Thus, the evolution of CDS spreads might be driven more by
liquidity constraints than by movements in the underlying sovereign risk
perception. Anyway, for a definitive validation, a scale effect conjecture
deserves further investigations, which are beyond the scopes of this paper.
\newline
Thus, this paper is organized as follows: Section \ref{experiments} contains
the description of the experiments, along with the presentation of the used
dataset. In Section \ref{experiments}, the main results of the experiment
are discussed. Section \ref{discussion} offers some concluding remarks
through a discussion of the results.

\section{Experiments}

\label{experiments}

In this section, we describe the empirical experiments carried out for
testing the conformity of the daily changes in CDS spreads with the BL.
First, the used dataset is introduced: for a better understanding of the
data scales, the main descriptive statistics are reported (see Subsection %
\ref{sub:data}). Furthermore, for the reader's convenience, we recall the
definition of CDS and their main properties. Second, the adopted
methodological tools are listed and discussed. The procedures implemented
for analyzing the deviations between the first digit of the CDS spreads and
BL distribution are also presented (see Subsection \ref{sub:meth}).

\subsection{Data}

\label{sub:data} We use daily Thomson Reuters Composite Sovereign CDS
spreads of 13 European countries provided by Data Stream, both inside and
outside the European Union and European Monetary Union. Thomson Reuters is
fairly acknowledged as one of the leading data providers of CDS quotes. The
Thomson Reuters end-of-day composite spreads are calculated by a standard
aggregation of the prices contributed by major market makers. With respect
to individual dealer's prices, the aggregated (composite) prices allow a
more comprehensive perspective on the market (Mayordomo et al., 2014).
\newline
We use daily CDS spreads in basis points\footnote{%
A basis point (bp) is equal to$1/100th$ of $1\%$.} (bps.) and two different
tenors (5 and 10 years), starting from August 2008 to April 2015. \newline
The time series considers 1,750 days, which results in a total of 42,000
spread observations. We recall that we consider four sets of countries, as
indicated here above : a) the \textit{core economies} - France Germany,
United Kingdom; b) the \textit{most worrying economies} - Ireland, Italy,
Portugal and Spain; c) the \textit{Eastern economies} - Croatia, Czech
Republic, Poland, Romania - and Turkey; d) Greece. The first two groups
contain countries on the basis of their credit risk: the first contains
countries with a credit rating of at least "A", while the second group
contains all CDS entities rated worse than "A". The third considers a
specific geographical area, Eastern economies and Turkey, while the fourth
contains only Greece, - which is the first European country that has
experienced in 2012 a partial default after the constituency of the currency
union. \newline
The time interval for such a study is taken as a whole: (i) August 8, 2008 -
April 25, 2015, but it is also useful to consider special subclasses: (ii)
August 8, 2008 - January 1, 2010; (iii) January 1, 2010 - October 31, 2013;
(iv) November 1, 2013 - April 25, 2015, and the (v) January 1, 2010 - April
25, 2015 which regroups the latter two time periods. The analysis was
conducted also on the sub-periods of the considered time span to highlight
whether the advent of the sovereign crisis may have been one cause of the
major/minor conformance of the daily changes in CDS quotes with BL. The
European sovereign debt crisis is, indeed, a multi-year crisis that has been
taking place through successive stages since the end of 2009. Especially
during the period (iii), it is common knowledge that the "most worrying
economies", faced a strong rise in spreads as the result of a widespread
concern about their future debt sustainability. \newline
During period (iv), European CDS quotes have shown, with respect to period
(iii), a downward trend in the CDS spread level, with the only exception of
Greece and Turkey. Table 2  (here inserted as Fig.2) reports some descriptive statistics and shows
that after the large upward jump observed during period (iii), the CDS
quotes of the core economies, together with worrying economies and eastern
economies, experienced a tendency to decrease. For instance, the mean 5
years German tenor decreased from 53.06 to 20.44 bps.; the corresponding
Italian one from 282.42 to 127.17 bps and the Portuguese from 601.71 to
204.92, only to mention some of them. It is worth to mention that a large
downward jump is observed also in the historical volatility (i.e. standard
deviation).

For reader's convenience, at this stage, it is fair to recall what a CDS is
aimed at and its general properties. Generally speaking, a CDS is an
agreement between two parties (the buyer and the seller). The buyer pays a
periodic premium, usually quarterly or semiannual, to hedge an underlying
security, a loan or a bond, against the default of the issuer. Upon the
default of the issuer, the seller commits herself/himself to pay the amount
the buyer will not recover from the default procedure. As a result of this
agreement, the buyer of a CDS receives credit protection, whereas its seller
guarantees the creditworthiness of the underlying debt security. So, the
risk of default is transferred from the holder of the fixed income
instrument to the seller of the swap. When CDS are used to hedge against the
default (or downgrade) of a State, they are called in jargon "\textit{%
sovereign CDS}". \newline
The Credit Default Swap (CDS), because of their above mentioned structure,
implicitly embed forward-looking information about the creditworthiness of
the issuer, -- given that CDS spreads are no more than the cost of credit
risk insurance. At the same time, CDS market is the source of regulatory
concerns due to its size and lack of transparency. As a matter of fact, the
CDS market is a decentralized unregulated OTC market in which detailed
information about the pricing mechanisms is rather scarce and contracts
often get traded so much that it is hard to know who stands at either end of
a transaction (Cont and Kokholm, 2014; Cont, 2010). \newline
Therefore, in order to investigate the (possible) presence of irregularities
in the CDS market more closely, we make use of the empirical Benford's
distribution, Eq. (1), and analyze daily changes in sovereign CDS quotes for
the mentioned European countries, inside and outside the European Union and
European Monetary Union, in time intervals admittedly characterized by the
effects of the sovereign debt crisis. We will find that CDS spreads depart
significantly from the expected Benford's reference distribution, raising
potential concerns relative to the unbiased nature of the warning signals
related with sovereign risk coming from the CDS market.

\subsection{Methodology}

\label{methodology}

\label{sub:meth}In the case of the daily changes in sovereign CDS quotes,
the statistical assessment of the closeness between the observed
distribution of digits and the corresponding values of BL is verified by
using 3 different approaches. First, the $\chi ^{2}$ test is computed on the
overall sample and on the sub-periods (ii)-(v) introduced in Subsection \ref%
{sub:data}. In doing so, we are able to statistically observe whether the
data fit the BL, at a given confidence level. Then, we compute the
Chebyshev's distance and the Kullback and Leibler's divergence in moving
windows, in order to track the time-dependence consistency of the observed
frequencies with the ones associated to BL. \newline
The $\chi ^{2}$ test is one of the most used statistical tools for checking
whenever a series of data satisfy a hypothesis, here the BL. In the first
digit case, there are nine possible outcomes -- ranging from 1 to 9 -- since
zero cannot be viewed as a significant first digit of a number. Thus, we
deal with $n=9$ values and $n-1=8$ degrees of freedom for the $\chi ^{2}$
test, according to the formula:
\begin{equation}
\chi ^{2}(8)=\sum_{i=1}^{9}\frac{\left( N_{obs}^{(i)}-N_{BL}^{(i)}\right)
^{2}}{N_{BL}^{(i)}},  \label{xi^2}
\end{equation}%
where $N_{obs}^{(i)}$ is the number of observations of the $i$-th digit and $%
N_{BL}^{(i)}$ is the related theoretical one suggested by the BL, for each $%
i=1,\dots ,9$. \newline
The critical values of the $\chi ^{2}$ test with 8 degrees of freedom and
the corresponding level of confidence are used to identify the acceptance
region. In the case of significance level of 0.05 and 8 degrees of freedom,
the critical value of the $\chi ^{2}$ test is 15.507. If the $\chi ^{2}$
test -- calculated as in (\ref{xi^2}) -- is below such a critical value,
then the null hypothesis of consistency of the data with the BL is accepted
at a significance level of 0.05. \newline
The Chebyshev's distance between two discrete probability distributions is
given as:
\begin{equation}
d_{C}(P_{obs},P_{BL})=\max\limits_{i=1,\dots ,9}|p_{obs}^{(i)}-p_{BL}^{(i)}|,
\label{Cheb}
\end{equation}%
where, -- in the peculiar case we deal with here, $P_{obs}=\left(
p_{obs}^{(1)},\dots ,p_{obs}^{(9)}\right) $ is the vector of the observed
frequencies for the first digit, while $P_{BL}=\left( p_{BL}^{(1)},\dots
,p_{BL}^{(9)}\right) $ is the one associated to BL. \newline
By adopting the same notation, the Kullback and Leibler's divergence between
the observed and BL frequencies is:
\begin{equation}
d_{KL}(P_{obs},P_{BL})=\sum_{i=1}^{9}p_{obs}^{(i)}\ln \left( \frac{%
p_{obs}^{(i)}}{p_{BL}^{(i)}}\right) .  \label{KL}
\end{equation}%
The interpretation of the Chebyshev's distance can be easily grasped when
looking at formula (\ref{Cheb}). It is easily argued that a high value of $%
d_{C}(P_{obs},P_{BL})$ indicates that there exists $i=1,\dots ,9$ such that $%
p_{obs}^{(i)}$ and $p_{BL}^{(i)}$ are remarkably different. \textit{A
contrario}, a low value of $d_{C}(P_{obs},P_{BL})$ is associated to small
deviations of $p_{obs}^{(i)}$ from $p_{BL}^{(i)}$, for each $i=1,\dots ,9$.
The extreme cases are $d_{C}(P_{obs},P_{BL})=0$ -- when $P_{obs}\equiv
P_{BL} $ in a component-wise sense -- and $d_{C}(P_{obs},P_{BL})=\max%
\limits_{i}|p_{BL}^{(i)}|=|p_{BL}^{(1)}|$ -- when $p_{obs}^{(1)}=0$. \newline
The Kullback and Leibler's divergence measures the information loss when $%
P_{obs}$ is taken as an approximation of the first digit distribution given
by $P_{BL}$. Then, $d_{KL}$ contributes to a correct evaluation of the
entity of the deviation of a set of data from BL when the first digit is
considered. Therefore, the Kullback and Leibler's divergence can be
interpreted as a proxy of the degree of opaqueness of the CDS market.
\newline
Next, we propose to examine the Chebyshev's distance and the Kullback and
Leibler's divergence in the view of a time track of the transparency of
sovereign CDS market. For this aim, 90-days rolling windows have been
considered, with a moving time of 45 days. The initial date is August 8,
2008 and the terminal one is April 25, 2015, whence leading to examine 38
time windows.

\section{Results}

\label{discussion} This section contains a critical reading of the results
of the analysis. We proceed by considering first the whole period along with
its four subperiods, and then a moving window approach.

\subsection{Whole time interval and its subperiods}

\label{sub1} Table 3  (here inserted as Fig.3) collects the values of $\chi ^{2}$ tests and related $p$%
-values for the considered 13 countries when the overall time interval and
the four sub-periods (ii)-(v) are taken into account.

Fixed a significance level of 0.05, Table 3   (inserted as Fig.3)  shows that -- when considering
the entire period -- the null hypothesis of BL is in general verified only
in the case of 10 yrs tenor and not in the 5 yrs one. This is true
especially for France (the $p$-value is about $0.95$) and Czech Republic ($%
0.97$), but also for Portugal ($0.89$), UK ($0.80$), Croatia ($0.80$), Spain
($0.78$), Ireland ($0.71$). This outcome suggests that 5 yrs CDS spreads are
probably the most "manipulated" ones. After all, it is useful to underline
that these discrepancies occur mainly in the commonly accepted reference
contract, being the 5 yrs tenor the most liquid in the CDS market. Greece
represents the only exception, with BL holding only for the changes in 5 yrs
CDS quotes with a $p$-value of $0.94$, decreasing to $0.75$ in the case of
10 yrs contracts. However, Greece is also a very peculiar case, not only
because of missing data, but also due to a persistent economic distress
which characterizes all the periods under investigation. BL is violated for
Germany, Poland, Italy, Romania and Turkey in both tenor cases. It is also
worth to remark here that the data of all the \textit{core economies}
present a remarkable discrepancy with BL in the case of 5 yrs tenor. These
findings again suggest that the most liquid contracts -- as the 5 yrs tenors
for the core economies -- are those more easily "manipulable". The analysis
of sub-periods provides more insights on when data show a higher deviation
from BL. The values of the $\chi ^{2}$ tests are higher (and BL is more
likely violated) over the sub-period 2010-2015 for all the considered
countries and tenors, with the only exception of the 4 Eastern European
Countries (Croatia, Czech Republic, Poland and Romania). However, focusing
on the sub-periods 2010-2013 and 2013-2015, it is possible to identify -- by
comparing the former period with the latter one -- a wider discrepancy with
BL for Germany, UK, Czech Republic (5 yrs tenor) and Germany, Italy,
Romania, Turkey (10 yrs tenor), a lower distortion for Italy, Croatia,
Poland, Romania, Turkey (5 yrs tenor) and France, Ireland, Poland, (10 yrs
case). Moreover, with the exception of Greece, the daily changes in 5 yrs
tenors CDS spreads never satisfy BL. In particular, splitting 2010-2015 in
two sub-periods (iii)-(iv) does not seem to produce any striking result
allowing us to identify a sub-period in which the violations of BL for core
economies occur, whence allowing us to conclude that such BL violations are
rather uniform and pervasive. Completely different is the outcome of the
analysis for the sub-period 2008-2010. In this case, the distribution of the
data is generally closer to BL, especially for the 5 yrs tenor case,
suggesting that data may have been objects of "manipulation" after 2010.

\subsection{Rolling windows}

\label{sub2} 
In order to better identify periods in which the distribution of sovereign 5
yrs CDS quotes deviates from BL, we performed an analysis on moving windows
to test the closeness of the distributions for rolling two months periods.
The tracking of the $\chi ^{2}$ tests by countries are reported in Table 4,  (here inserted as Fig.4) 
where the values of the test by country and moving window are reported.

\begin{figure}
 \begin{center} 
 \includegraphics  [height=22.6cm,width=18.8cm]  {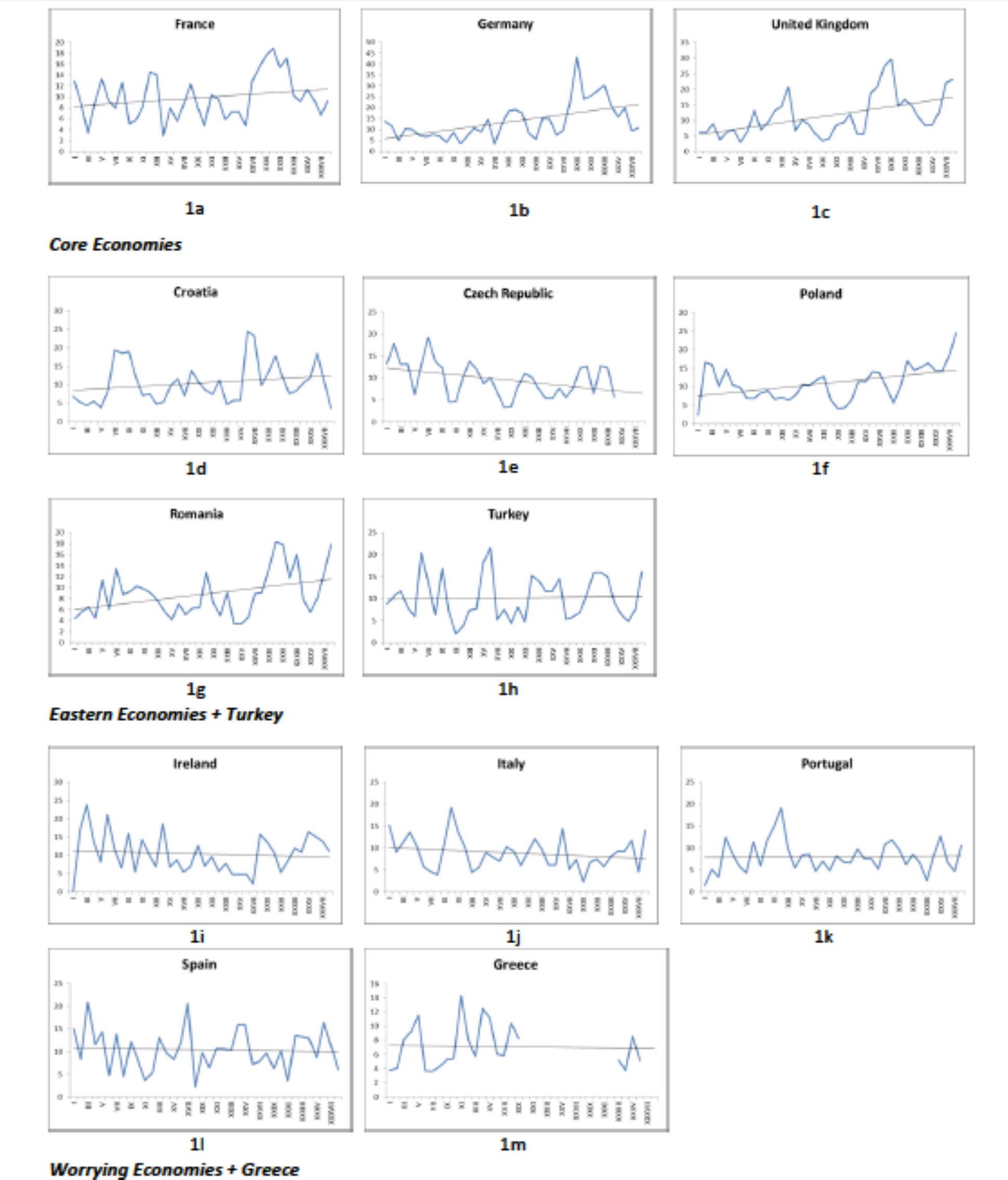}
\caption{  Linear trends for the $\chi^2$ tests over the moving
windows  }
\label{Figure1}
 \end{center}
\end{figure}

Considering a significance level of 0.05, an overview of the results
suggests mixing evidence. Some windows are associated to a great departure
from the BL (see e.g. the high values of the $\chi ^{2}$ test over the
windows XXVIII-XXXIII for the core economies, over the windows XXVI-XXVII
for Croatia, XXXI-XXXVII for Poland) while in other ones there is
concordance with BL (see e.g. window III for France, XII for Germany, IV for
UK, VIII for Italy and the very low value of $\chi ^{2}$ test -- i.e.: 1.49
-- in the first window for Portugal). It seems that the distance between the
distributions tends to increase with respect to time, as is evidenced by
Figure 1, graphics a-c, where a trend line is superimposed on the temporal
evolution of the values of the $\chi ^{2}$ test. With the Czech Republic and
Turkey exceptions, the group Eastern economies + Turkey shows the same
tendency, that is the distance between the two distributions increases when
one gets closer to the advent of the sovereign debt crisis (see Table 4  (inserted as Fig.4)  and
Figure 1, d-h). Similar results are obtained in the case of countries which
are in the group of the worrying economies (see Table 4  (inserted as Fig.4) and Figure 1, i-m).
In summary, contrary to what happens for the group of worrying economies, it
seems that CDS quotes of core economies have been "manipulated", with
increasing evidence during the European debt crisis. \newline
Finally, Tables 5 and 6  (here inserted as Figs. 5 and 6) show respectively the values of the Chebyshev's distance and Kullback and Leibler's divergence by moving window and country.

The tracking of the Chebyshev's distance (see Table 5  (here inserted as Fig. 5)) and the Kullback and
Leibler's divergence (see Table 6  (inserted as Fig. 6)), confirm some of the above mentioned
results (for the discordance case see e.g., in Table 5  (here inserted as Fig. 5), the high values in
windows XXVIII-XXX for the core economies, in window XXVI for Croatia,
XXXI-XXXII for Poland and, for Table 6  (inserted as Fig. 6), windows XXVIII-XXXI for Germany and
UK, windows XXVI-XXVII for Croatia and XXXI-XXXIII for Poland; for the
concordance with BL see e.g. window I for Portugal and window III for France
in both Tables). Furthermore, Table 5 and Table 6 (here inserted as Fig. 5 and Fig. 6) allow us to add some
conclusion to previous results. This is possible because the Chebyshev's
distance captures the largest deviation between the empirical frequencies of
the numerical values of the first digits and those associated to BL, while
the Kullback and Leibler's divergence represents a proxy of the information
loss due to the approximation of the BL through the empirical frequencies of
the first digits. In this respect, Table 5 (here inserted as Fig. 5) shows that the Chebyshev's
distance ranges over a rather narrow interval. However, it seems to assume
higher values in the most recent rolling windows for some core countries
(Germany and UK), exhibits a U-shape form in the case of Italy and Poland
while it is a reversed U-shape in the remarkable case of Greece (but notice
the large amount of missing data). In terms of levels, it is important to
note that Germany, UK, Czech Republic and Poland exhibit very high maximum
values of the Chebyshev's distance (more than 0.18). This evidence suggests
that in some rolling windows BL is violated because 1 -- the most frequent
first digit -- appears less than BL would suggest. The countries whose
sovereign CDS exhibit the smallest maximum deviation from BL are Portugal
and Turkey. \newline
Looking at Table 6 (here inserted as Fig. 6), it can be observed that there exists a positive trend
for UK, Poland, Romania and Turkey. For these countries, the information
loss in explaining BL through the empirical distributions is very large for
the most recent rolling windows. A few countries (UK, Czech Republic and
Poland) show a rather high value of the maximum of the Kullback and
Leibler's divergence (more than 0.27, with the peak of UK at 0.37),
suggesting a higher level of deviation over some sub-periods. Differently,
Italy and Portugal seem to exhibit less distorted distribution.

\section{Conclusions}

In this article, we adopt a data science perspective and discuss the
validity of the BL for some European sovereign CDS over a eight years
period, which includes the current financial crisis. We have dealt with this
issue by adopting different points of view. Firstly, a global analysis over
the entire period on the fitting of BL with the daily changes in CDS for the
first digits has been performed. Secondly, further elaborations have been
performed by clustering the reference period into four meaningful
sub-periods. Moreover, a tracking study has been carried out by considering
a moving window approach. We illustrated the usefulness of BL to study the
data quality of sovereign CDS. The employed statistical tools have been the $%
\chi ^{2}$ test -- which provides a general goodness-of-fit measure, the
Chebyshev's distance -- leading to results on the specific changing of the
most manipulated digit value and the Kullback and Leibler's divergence which
is a measure of the loss of information in passing by BL to empirical
frequencies. \newline
Results suggest that the most liquid contracts seem to be the most
"manipulated" ones, and deviations involve mainly the core economies and the
5 yrs tenor case. It is worth to note that Greece follows a very different
path with respect to the other European countries. Furthermore,
"manipulations" seem to have occurred mainly after the beginning of the
sovereign debt crisis.

  \begin{figure}
 \includegraphics  [height=26.8cm,width=20.2cm] 
  {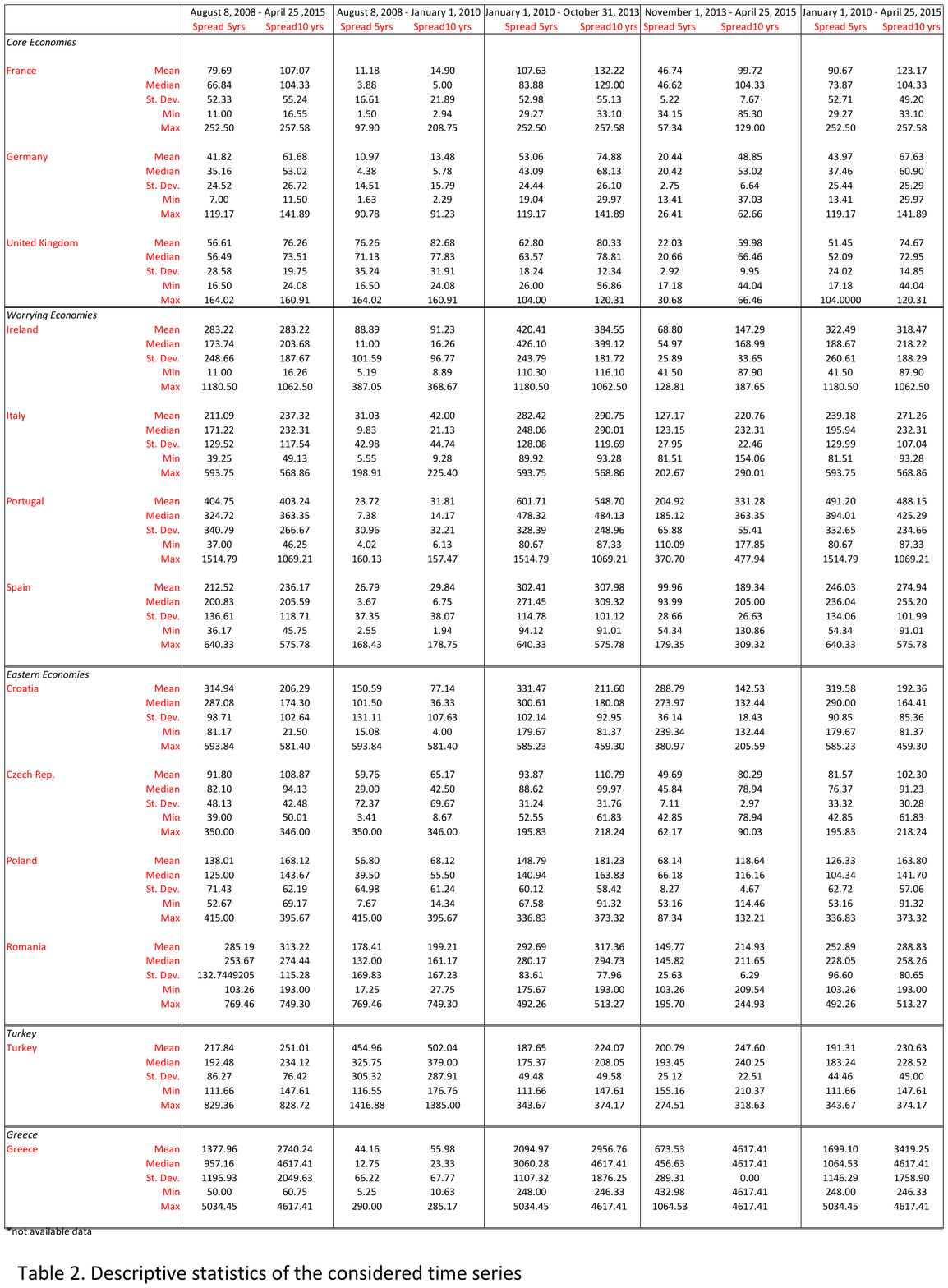}
 \caption{  
 }
\label{Table2}
\end{figure}

 \begin{figure}
 \includegraphics [height=34.8cm,width=20.2cm] 
  {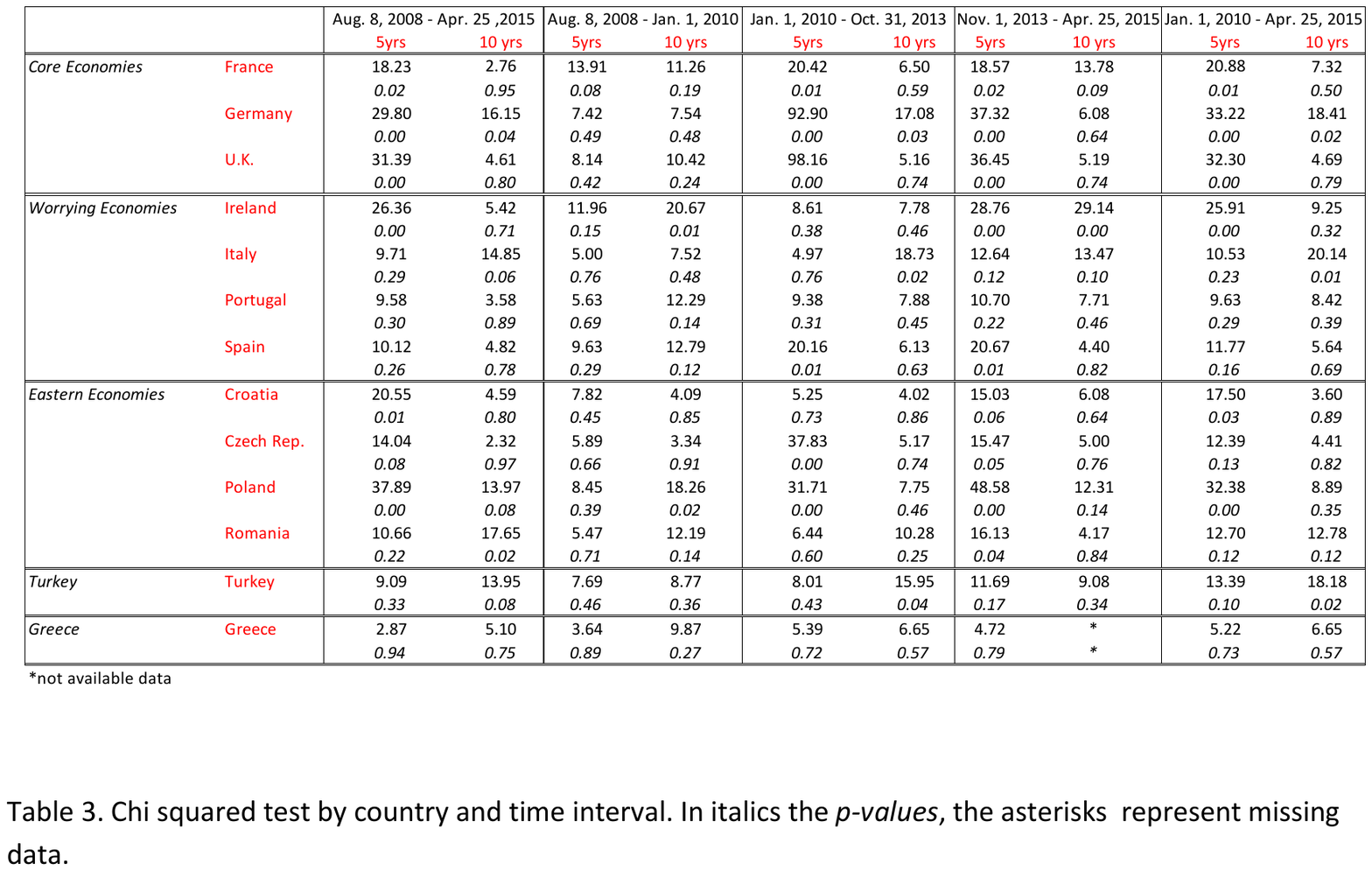}
\caption{ Table  3
}
\label{Table 3}
\end{figure}

 \begin{figure}
 \includegraphics [height=26.8cm,width=18.8cm]  
 {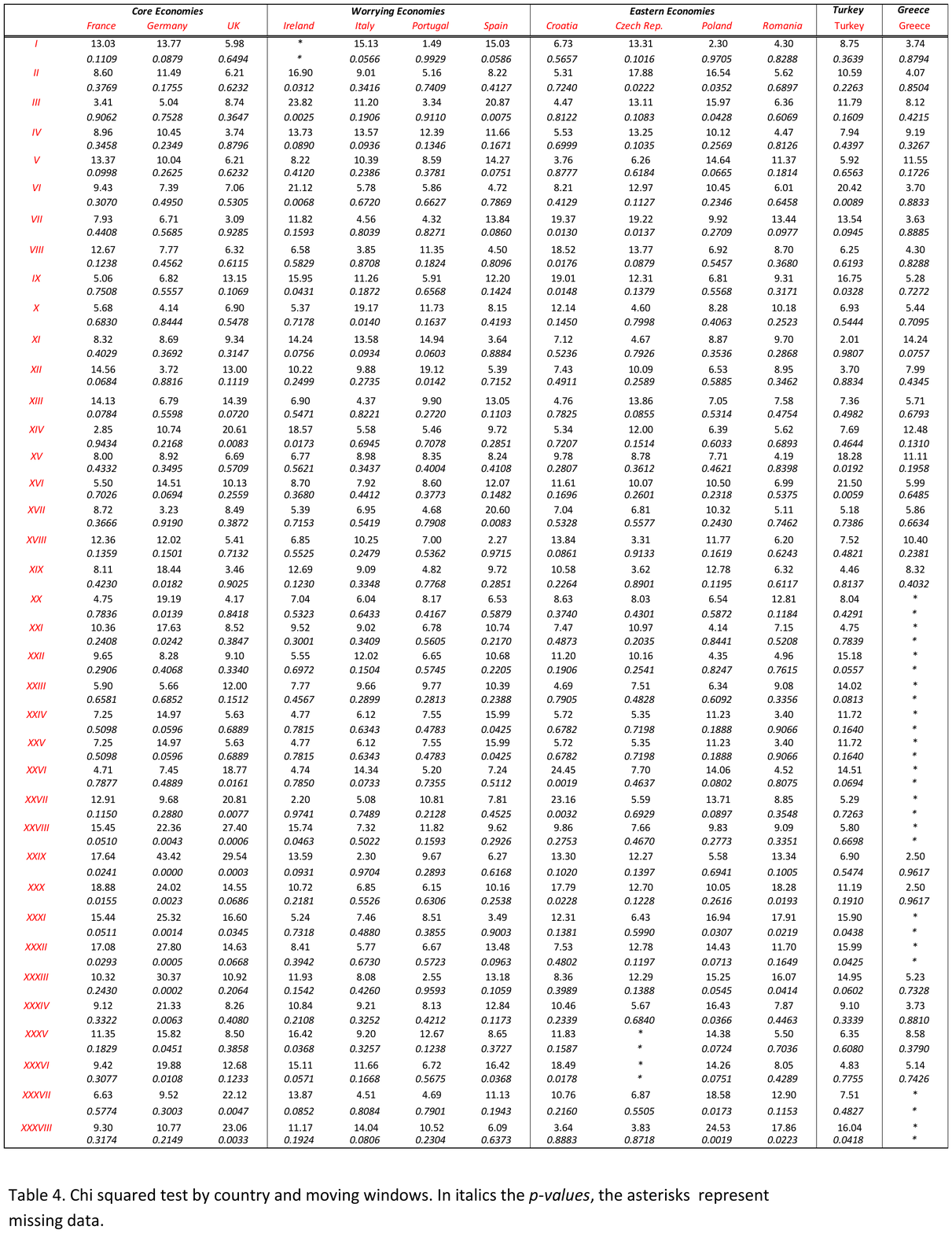}
\caption{ Table  4: 
}
\label{Table 4}
\end{figure}

 \begin{figure}
 \includegraphics [height=24.8cm,width=18.8cm]  
 {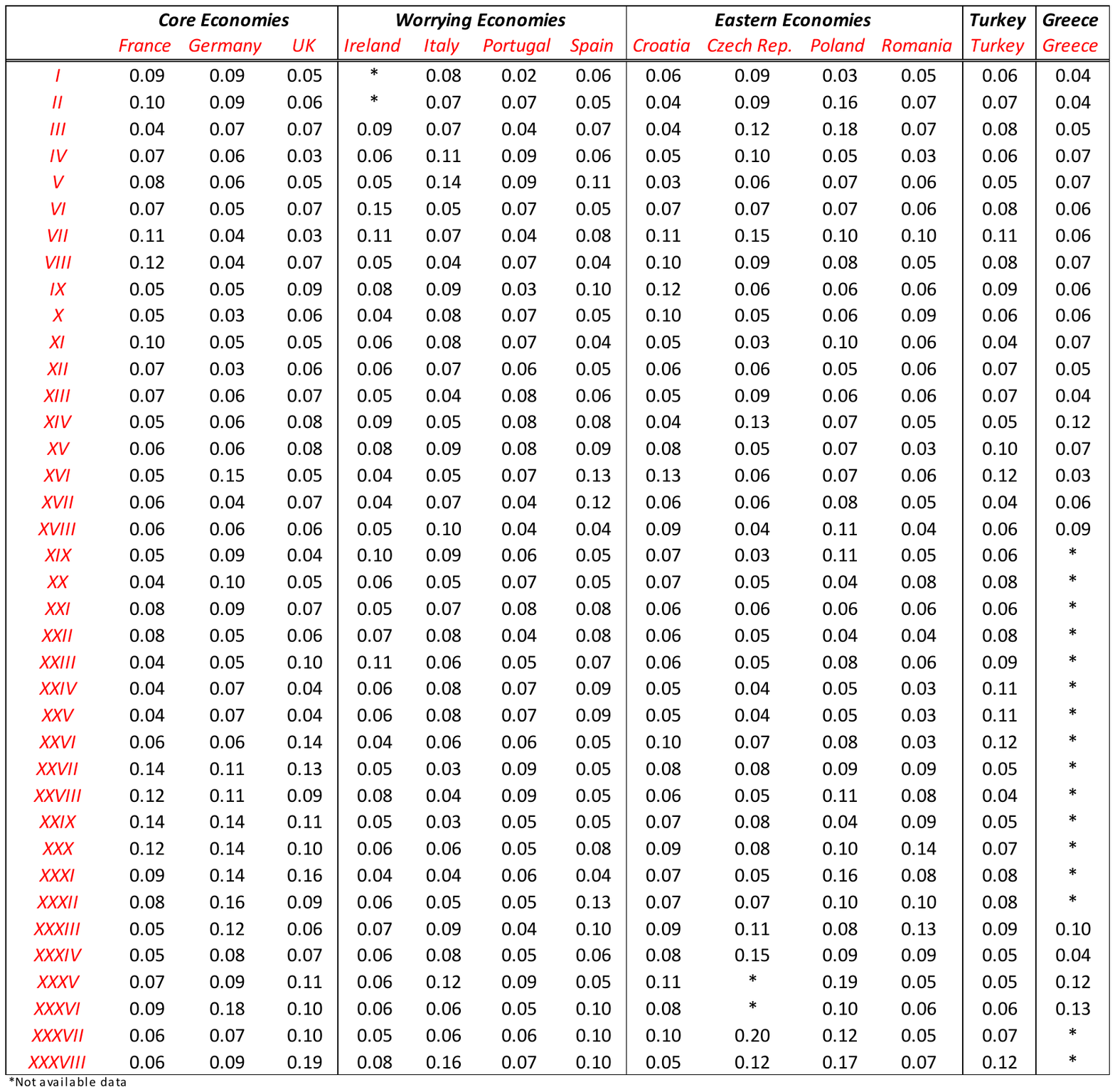}
\caption{  Table  5 : Chebyshev's distance by moving windows and country.
}
\label{Table 5}
\end{figure}

 \begin{figure}
 \includegraphics [height=24.8cm,width=18.8cm] 
 {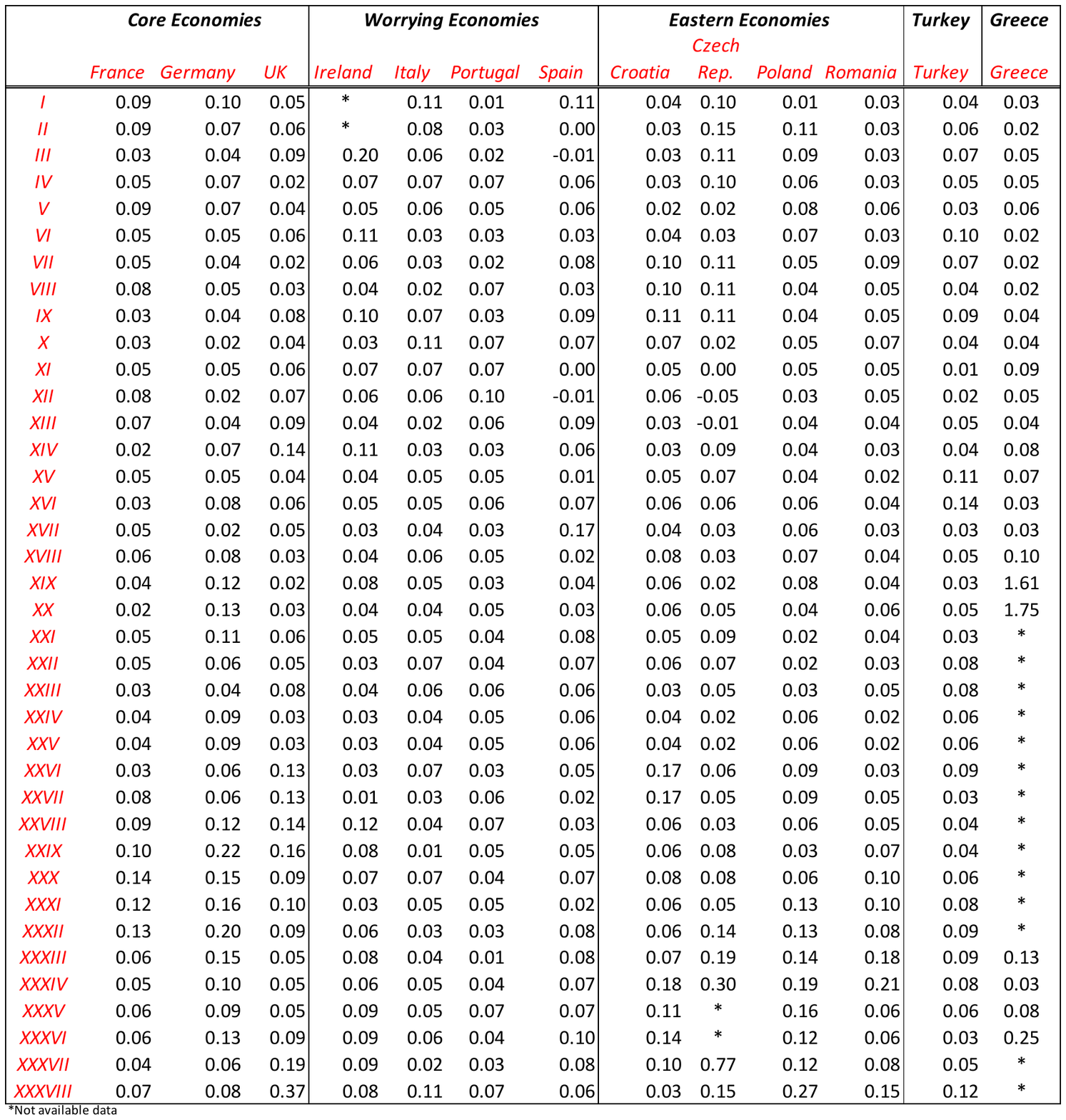}
\caption{  Table  6 :  Kullback and Leibler's divergence by moving window and country.
}
\label{Table 6}
\end{figure}



\end{document}